\documentclass[floats,preprint,prl,aps]{revtex4}
\usepackage{latexsym}
\usepackage{graphicx}
\usepackage{graphicx}
\usepackage{epsfig,amssymb}
\newcommand\beq{\begin{equation}}
\newcommand\eeq{\end{equation}}
\newcommand\bea{\begin{eqnarray}}
\newcommand\eea{\end{eqnarray}}
\newcommand\non{\nonumber}
\newcommand\bib{\bibitem}

\begin{document}
\title{\bf Topological quantum phase transition of light 
}
\author{\bf Chandan GN, N. Banerjee and Sujit Sarkar}
\address{\it Poornaprajna Institute of Scientific Research,
4 Sadashivanagar, Bangalore 5600 80, India.\\
e-mail: sujit.tifr@gmail.com \\
}
\date{\today}

\begin{abstract}
We study theoretically the topological quantum phase transition in 
Cavity QED lattice. We predict the condition for 
non-topological phase to the topological phase 
transition conditions for three different model Hamiltonians in cavity QED lattice. 
We study these topological quantum phase transition through winding number, 
which is a topological 
invariant quantity.
We argue that the appearance of topological phase in these 
systems where the discrete $Z_2$ symmetry broken.
We show that the non-topological state is the vacuum state of the
system where each cavity contains fermionic type excitations from 
light-matter interaction whereas the topological state of system 
contains Majorana modes of excitations at the end cavity of the lattice. \\

\end{abstract}
\maketitle

\section{I. Introduction}
In condensed
matter system, quantum phase transition plays a significant role to
study and explain different quantum phases of the system which explain the different
broken symmetry states of the system, which are used to describe with the
concept of local order parameter of Ginzburg-Landau Theory \cite{berni,ss}.\\
The major limitation of Ginzburg-Landau theory is that the order parameter is treated 
as a local order parameter, this is over come by 
Topological quantum phase transition \cite{berni}.
In recent years, it reveals experimentally and theoretically in condensed matter
system that the typical order parameter is not a local order parameter rather it
is highly non-local order parameter \cite{berni}. 
One of the example for the non local order parameter is quantum
Hall state which corresponds to the annihilating an electron at a position by 
unwinding the number of fluxes.
This flux unwinding process is highly non-local and Ginzburg-Landau 
theory is not able to describe this phenomena.
The states with non-local order parameter are termed topologically ordered. 
The fact that topology is a characterization of global shape and 
it is invariant under a small local deformation it becomes one of the 
important property of the system [1]. 
This topological robustness protects the quantum state from the external perturbation. 
As a consequence of it several interesting physical phenomena 
appears in the low dimensional quantum many body system \cite{berni,nayak}. \\ 
As the physics of spontaneous symmetry breaking is absent 
for the topological state of matter, 
the concept of order parameter is also absent to explain the relevant 
low energy physics of these systems. 
The topological phases are characterized by topological 
invariants integer number.  
In the present research problem, we do the explicit study of the
topological quantum phase transition through winding number study.\\ 
{\bf Motivation:} \\
The study of topological phases of quantum many body systems are still in the beginning phase. 
Therefore to search the topological states in different physical system is increasing rapidly. 
The topological phases of condensed matter system attract much more attention due to 
their practical application in low dimensional quantum many body system [1,3]. 
Here we state few examples of topological states of matter in condensed matter physics.\\
Integer quantum Hall effect, fractional quantum Hall effect, 
quantum spin Hall effect, topological insulator. But the topological state of matter
for cavity QED has not explored yet to the mark \cite{horo,agarwal}.\\ 
The other part of the motivation comes from the Kitaev's seminal paper \cite{kitaev}. 
Kitaev has proposed the model 
of one dimensional spinless p-wave superconductors to 
realize the existence of topological phase. 
At that period, it is difficult to realize Kitaev's model in reality. 
The electron carry spin-1/2, the first step is to 
freeze the spin of the particle so that the system
appears as a one dimensional spinless system. 
In the interacting light-matter system,
specially in the cavity QED system where the experimental advancement is in the state
of art and the spin of the system mimics the state level difference
and the measurement of quantum state is extremely precise \cite{horo,agarwal}. 
In low-dimensional interacting light-matter system excitations appear as a collective mode  
under certain
physical conditions it behaves as a Majorana fermion mode \cite{sujop}.
Therefore we decide to explain the topological state of interacting light-matter
system and also to realize the Kitaev's model for cavity QED system.\\
\section{ II. The Model Hamiltonians and The Derivation of Effective Hamiltonians}
The Hamiltonian of our present study consists of three parts:
\beq
H ~= ~ {H_A} ~+~ {H_C}~+~{H_{AC}}
\eeq     
The Hamiltonians are the following
\beq
{H_A} ~=~ \sum_{j=1}^{N} { {\omega}_e } |e_j > <e_j | ~+~ 
{\omega}_{ab} |b_j > <b_j | , 
\eeq
where $j$ is the cavity index. ${\omega}_{ab} $ and ${\omega}_{e} $ are 
the energies of the state $ | b> $ and the excited state respectively. The
energy level of state $ |a > $ is set as zero. $|a>$ and $|b> $ are
the two stable state of a atom in the cavity and $|e> $ is the
excited state of that atom in the same cavity. 
The following Hamiltonian describes the photons in the cavity,
\beq
 {H_C} ~=~ {{\omega}_C} \sum_{j=1}^{N} {{a_j}}^{\dagger} {a_j} ~+~
{J_C} \sum_{j=1}^{N} ({{a_j}}^{\dagger} {a_{j+1}} + h.c ),  
\eeq 
where ${a_j}^{\dagger}({a_j})$ is the photon
creation (annihilation) operator for the photon field in the $j $'th cavity, ${\omega}_C $
is the energy of photons and $ J_C $ is the tunneling rate of photons
between neighboring cavities.
The interaction between the atoms and photons and also by the driving lasers
are described by
\beq
{H_{AC}}~=~ \sum_{j=1}^{N} [ (\frac{{\Omega}_a}{2} e^{-i {{\omega}_a} t} +
{g_a} {a_j}) |e_j > < a_j | + h.c] + [a \leftrightarrow b ] . 
\eeq
Here ${g_a} $ and ${g_b} $ are the couplings of the cavity mode for the
transition from the energy states $ |a > $ and $ | b> $ to the excited state.
${\Omega}_a $ and ${\Omega}_b $ are the Rabi frequencies of the lasers
with frequencies ${\omega}_a $ and $ {\omega}_b $ respectively.\\
The authors of Ref. \cite{hart1}
have derived an effective spin model by considering the following physical
processes:
A virtual process regarding emission and absorption of
photons between the two stable  states of neighboring cavity yields the resulting 
effective Hamiltonian as
\beq	
{H_{xy}} = \sum_{j=1}^{N}  B {{\sigma}_j}^{z} ~+~\sum_{j=1}^{N} 
(\frac{J_1}{2} {{\sigma}_j}^{+} {{\sigma}_{j+1}}^{-} ~+~
\frac{J_2}{2} {{\sigma}_j}^{-} {{\sigma}_{j+1}}^{-} + h.c ).  
\eeq 
When $J_2 $ is real then this Hamiltonian reduces to the XY model.
Where ${{\sigma}_j}^{z} = |b_j > <b_j | ~-~ |a_j > <a_j | $,
${{\sigma}_j}^{+} = |b_j > <a_j | $, ${{\sigma}_j}^{-} = |a_j > <b_j | $
. 
\bea
H_{xy} & = & \sum_{i=1}^{N} ( B  {{\sigma}_i}^{z}~+~ {J_1} ( {{\sigma}_i}^{x}
{{\sigma}_{i+1} }^{x}  + {{\sigma}_i}^{y}
{{\sigma}_{i+1} }^{y} ) \non\\
& & + {J_2} ( {{\sigma}_i}^{x}
{{\sigma}_{i+1} }^{x} - {{\sigma}_i}^{y}
{{\sigma}_{i+1} }^{y} ) ) \non\\    
& & =  \sum_{i=1}^{N} B ( {{\sigma}_i}^{z}~+~{J_x} {{\sigma}_i}^{x}
{{\sigma}_{i+1}}^{x} ~+~ {J_y} {{\sigma}_i}^{y}
{{\sigma}_{i+1}}^{y}) .
\eea
With ${J_x} = (J_1 + J_2 ) $ and ${J_y} = (J_1 - J_2 ) $.\\
We follow the references \cite{hart1}, to present the analytical 
expression for the different physical parameters of the system.\\
\beq
B = \frac{\delta_1}{2} - \beta 
\eeq
\beq
{J_1} = \frac{\gamma_2}{4} ( \frac{{|{\Omega_a}|}^2 {g_b}^2 }{{ {\Delta}_a }^2 }
 +  \frac{{|{\Omega_b}|}^2 {g_a}^2 }{{ {\Delta}_b }^2 } ) ,
{J_2} = \frac{\gamma_2}{2} ( \frac{{\Omega_a} {\Omega_b} g_a g_b }{{\Delta}_a {\Delta_b} }
 ).
\eeq
The detail analytical expression for ${\gamma}_{ab} , {\gamma}_1 , {\gamma}_2,
{\delta}_1, {\Delta}_a , {\Delta}_b, {\delta_a}^{k} , {\delta_b}^{k}$ and
$\Omega_k $ are relegated to the appendix. \\ 
Here we discuss very briefly about an effective $z$-component of
interactions
($ {{\sigma}_i}^{z} {{\sigma}_{i+1}}^{z}$) in such a system. The authors of
Ref.\cite{hart1,hart2}
have proposed the same atomic level configuration but having only one
laser of frequency ${\omega}$ that mediates the atom-atom coupling through
virtual photons. Another laser field with frequency $\nu $ is used to
tune the effective magnetic field.
In this case the Hamiltonian ${H_{AC}} $ changes but the Hamiltonians $H_A $
and $H_C $ are the same.
\bea
{H_{AC}}&=& \sum_{j=1}^{N} [ (\frac{{\Omega}}{2} e^{-i {{\omega}}t} +
\frac{{\Lambda}}{2} e^{-i {{\nu}_a}t}
{g_a} {a_j}) |e_j > < a_j | + h.c] \non\\
& & + [a \leftrightarrow b ] .
\eea
Here, ${\Omega}_a $ and ${\Omega}_b $ are the Rabi frequencies of the
driving laser with frequency ${\omega}$  on transition $|a > \rightarrow |e> $
, $|b > \rightarrow |e> $, whereas ${\Lambda}_a $ and ${\Lambda}_b $ are the
driving laser with frequency ${\nu}$  on transition $|a > \rightarrow |e> $
, $|b > \rightarrow |e> $. One can eliminate adiabatically the excited atomic
levels and
photons by considering the interaction picture with respect to
$ H_0 = H_A ~+~H_C $ [6,7]. They have considered the detuning parameter in such
a way that the Raman transitions between two level are suppressed and also
chosen the parameter in such a way that the dominant two-photon processes are
thus no transition between levels a and b. Whenever two atoms exchange a
virtual photon both of them experience a Stark shift and play the
role of an effective $ {{\sigma}^{z}}{{\sigma}^{z}} $ interaction
\cite{hart1,hart2}. Then
the effective Hamiltonian reduces to
\beq
{H_{zz}}~=~\sum_{j=1}^{N} ( {B_z} {{\sigma}_j}^{z} ~+~ {J_z}
{{\sigma}_j}^{z}{{\sigma}_{j+1}}^{z} ) .
\eeq
These two parameters can be tuned independently by varying the laser frequencies.
Finally, they have obtained an effective model by combining Hamiltonians $H_{xy} $ and
$H_{zz} $ by using Suzuki-Trotter formalism \cite{hart1,hart2}. 
The effective Hamiltonian
simulated by this procedure is
\beq
H_{spin} ~=~\sum_{j=1}^{N} ( B_{tot} {{\sigma}_j}^{z} ~+~
\sum_{{\alpha}=x,y,z} J_{\alpha} {{\sigma}_j}^{\alpha} {{\sigma}_{j+1}}^{\alpha}) ,
\eeq
where $ B_{tot} = B + {B_z} $. 
It has been shown in Ref. \cite{hart2}
that $J_y $ is less than
$J_x $. From the analytical expressions of
$J_x $ and $ J_y $, it is clear that the magnitudes of ${J_1}$ and $J_2 $
are different.
$ {J_z}= {\gamma}_2 {| \frac{ {{\Omega}_b}^{*} g_b }{4 \Delta_b}
- \frac{ {{\Omega}_a}^{*} g_a }{4 \Delta_a} |}^2 $.\\
$ B_{tot} = -\frac{1}{2} [ \frac{ {|{\Lambda}_b |}^2 }{16 {\tilde{\Delta_b}}^2}
( 4 \tilde{\Delta_b } -  \frac{{|{\Lambda_a}|}^2 }{4 ( {\tilde {\Delta}_a}  -
{\tilde {\Delta}_b} )} -  \frac{{|{\Lambda_b}|}^2 }{{\tilde {\Delta}_b } }
- {\beta_2 }) - {\beta_3}]. $ \\
The quantum state engineering of cavity QED is in the state of art due to the
rapid progress of technological development of this field [3,4]. Therefore one can 
achieve this limit to get the desire Hamiltonian and 
quantum state of the system, when we consider the situation
where $ J_y =0 $ and $ J_z =0 $. In this limit the atom-photon coupling strength
$g_a = g_b$. The detail derivation is relegated to the appendix. \\
From the above equation, We get the following relations, $ g_b = g_a $ to get
the transverse Ising model. The detail derivation is relegated to the appendix. 
\\
\beq
H_{1} = \sum_{j=1}^N (B {{\sigma}_z} (j) + J_x {{\sigma}_x} (j) {{\sigma}_{x}} (j+1) ).
\eeq
One can write the above Hamiltonian in spinless fermion operators by using the Jordan-Wigner
transformation. To do so, we use the following relation. 
\bea
{\sigma_n}^{x} {\sigma_{n+1}}^{x} & = & ({{\psi}_n }^{\dagger} - {\psi_n}) 
({{\psi}_{n+1} }^{\dagger} + {\psi_{n+1} }) \non\\ 
{\sigma_n}^{y} {\sigma_{n+1}}^{y} & = & ({{\psi}_n }^{\dagger} - {\psi_n}) 
({{\psi}_{n+1} }^{\dagger} - {\psi_{n+1} }) \non\\ 
{\sigma_n}^{z} {\sigma_{n+1}}^{z} & = & (2 {{\psi}_n }^{\dagger}{\psi_n} -1 ) 
(2 {{\psi}_{n+1} }^{\dagger}{\psi_{n+1} } -1 )  
\eea
The Hamiltonian, $H_1 $ become,
\beq
H_{1} = J_x \sum_{n} ( {\psi}^{\dagger} (n) {\psi} (n +1 ) + h.c )
+ J_x \sum_{n} ( {\psi}^{\dagger} (n) {\psi}^{\dagger} (n +1 ) + h.c )
+ 2 B \sum_{n} {\psi}^{\dagger} (n) {\psi} (n) .
\eeq
We get this Hamiltonian for the condition $g_a = g_b$. \\
Similarly for the Hamiltonian, $H_2 $, where $J_x$ and $J_y$ are
non-zero. One can write the Hamiltonian in the following form.\\
\beq
H_{2} = ( J_x + J_y) \sum_{n} ( {\psi}^{\dagger} (n) {\psi} (n +1 ) + h.c )
+ | J_x - J_y| \sum_{n} ( {\psi}^{\dagger} (n) {\psi}^{\dagger} (n +1 ) + h.c )
+ 2 B \sum_{n} {\psi}^{\dagger} (n) {\psi} (n) .
\eeq
We get this Hamiltonian for the condition, 
$ {{\Omega}_b}^{*} g_b {\Delta}_a = {{\Omega}_a}^{*} g_a {\Delta}_b $. 
The detail derivation is relegated to the appendix.\\
Similarly for the Hamiltonian, $H_3 $, where $J_x$, $J_y $ and $J_z$ are 
non-zero. \\
\beq
H_{3} = ( J_x + J_y) \sum_{n} ( {\psi}^{\dagger} (n) {\psi} (n +1 ) + h.c )
+ | J_x - J_y| \sum_{n} ( {\psi}^{\dagger} (n) {\psi}^{\dagger} (n +1 ) + h.c ) \\
+ (2 B + 8 \rho -4) \sum_{n} {\psi}^{\dagger} (n) {\psi} (n) .
\eeq
where $\rho = <{{\psi}_n}^{\dagger} {\psi}_n > $ is the density of excitation in
interacting light-matter system. Here we do the many body physics 
decoupling scheme to reduce the quartic interaction of the Hamiltonian to quadratic
one. 
\beq
{\sigma_n}^{z} {\sigma_{n+1}}^{z} = ( 8 \rho - 4) {\psi}^{\dagger} (n) {\psi} (n)   
\eeq
After the Fourier transformation, the Hamiltonian, $H_1$ , reduce to,
\bea
H_1  & = & 2 \sum_{k> 0} (2 B + J_x  cosk) 
({c_k}^{\dagger} {c_k} + {c_{-k}}^{\dagger} {c_{-k}}) \non\\
& & + 2 i J_x  \sum_{k > 0} sink  ({c_k}^{\dagger} {c_{-k}}^{\dagger} + 
{c_{k}} {c_{-k}}) .
\eea  
Similarly for the Hamiltonian, $H_2 $ reduced to, 
\\
\bea
H_2  & = & 2 \sum_{k> 0} ( 2 B + ( J_x + J_y)  cosk) 
({c_k}^{\dagger} {c_k} + {c_{-k}}^{\dagger} {c_{-k}}) \non\\
& & + 2 i | J_x - J_y|  \sum_{k > 0} sink  ({c_k}^{\dagger} {c_{-k}}^{\dagger} + 
{c_{k}} {c_{-k}}) .
\eea  
Similarly the Hamiltonian, $H_3 $ reduced to,
\\
\bea
H_3  & = & 2 \sum_{k> 0} ( (2 B + 8 \rho -4) + ( J_x + J_y)  cosk) 
({c_k}^{\dagger} {c_k} + {c_{-k}}^{\dagger} {c_{-k}}) \non\\
& & + 2 i | J_x - J_y|  \sum_{k > 0} sink  ({c_k}^{\dagger} {c_{-k}}^{\dagger} + 
{c_{k}} {c_{-k}}) .
\eea 
Now our main interest is to study the topological quantum phase transition in cavity
QED lattice system based on these models. 
Our starting point is to recast our three Hamiltonians
($ H_1 , H_2 , H_3 $)
of interacting light-matter system.
 \\
\beq 
H_{J=1,2,3} = [ \sum_{n} -{t}^{J}  ( {c_n}^{\dagger} c_{n+1} + h.c ) -
 {\mu}^{J} {c_n}^{\dagger} {c_n} 
-  {|\Delta|}^{J} ( {c_n} {c_{n+1}} + h.c ) ] .
\eeq
In the above Hamiltonian, we neglect the common negative sign which
will not alter the relevant physics of the system.\\ 
$ {t}^{(1)} = J_{x} $, $ {\Delta}^{(1)} = J_{x} $ and ${\mu}^{(1)} = 2 B$. \\
$ {t}^{(2)} = ( J_{x} + J_{y}) $, $ {\Delta}^{(2)} = ( J_{x} - J_{y}) $ and 
${\mu}^{(2)} = 2 B$. \\
$ {t}^{(3)} = ( J_{x} + J_{y}) $, $ {\Delta}^{(3)} = ( J_{x} - J_{y}) 
$ and ${\mu}^{(2)} = 2 B + 8 \rho -4 $. \\
The bulk properties of Hamiltonian can be studied in the momentum space. One
can write down the Hamiltonian in momentum space as.\\
\beq 
H^k_{J=1,2,3} = (\frac{1}{2}) \sum_{k} {\psi_k}^{\dagger} {H^k_{J=1,2,3}} {\psi_k} 
\eeq
$ {H}_{J=1,2,3} = \left (\begin{array}{cc}
      {\epsilon}^{(J=1,2,3)} (k)  & 2 {{\Delta}^{(J=1,2,3)}}^{*} (k) \\
   2 {{\Delta}^{(J=1,2,3)}} (k)   & - {\epsilon}^{(J=1,2,3)} 
        \end{array} \right ) $
where, $ {\epsilon}^{(J=1,2,3)} = -2 t^{(J=1,2,3)} cosk - {\mu}^{(J=1,2,3)} $,
and $ {\Delta^{(J=1,2,3)}}(k) = - i {\Delta}^{(J=1,2,3)} sink $.\\
This Hamiltonian corresponds to the p-wave superconducting
phase, one can understand this in the following way. \\
One can also write down the above Hamiltonian in Bogoluibov
energy spectrum,
\beq
H_{J=1,2,3} = \sum_{k}  { E(k)}^{J} {a_k}^{\dagger} {a_k} 
\eeq 
Here $E_k $ is the energy spectrum in bulk and ${a_k}^{\dagger} $ and $a_k $ 
are the Bogoliubov quasiparticles operators.
\begin{figure}
\includegraphics[scale=0.4,angle=0]{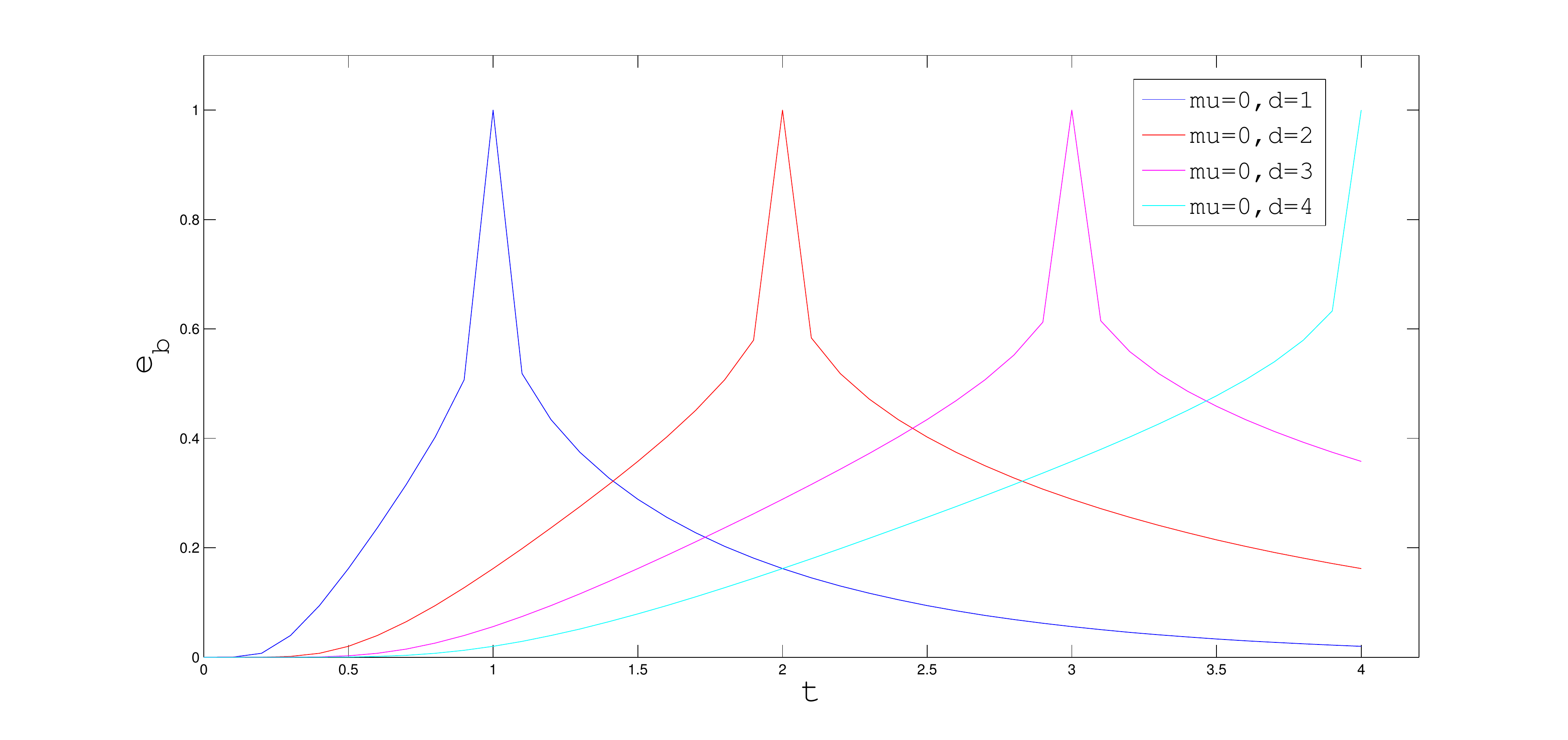}
\includegraphics[scale=0.4,angle=0]{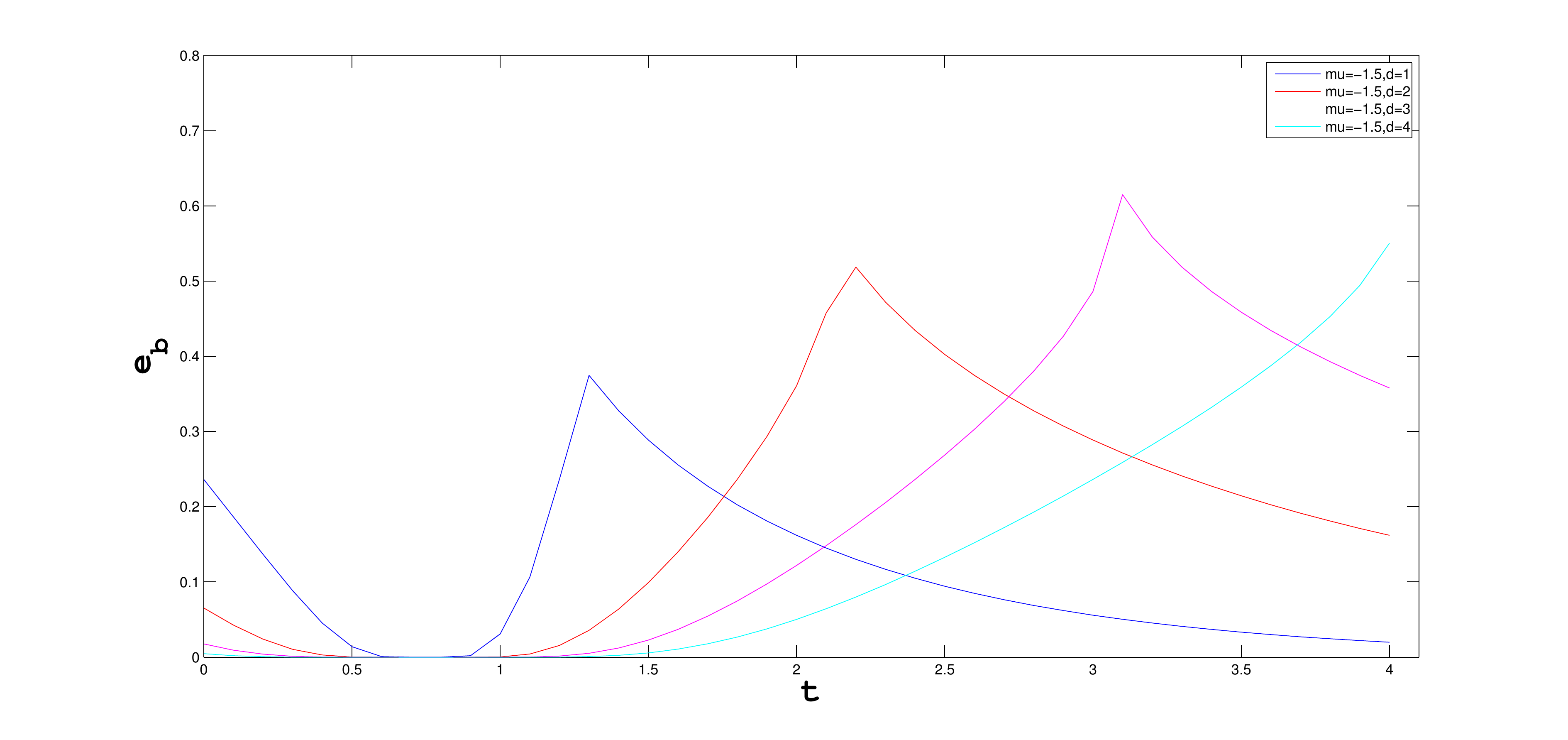}
\includegraphics[scale=0.4,angle=0]{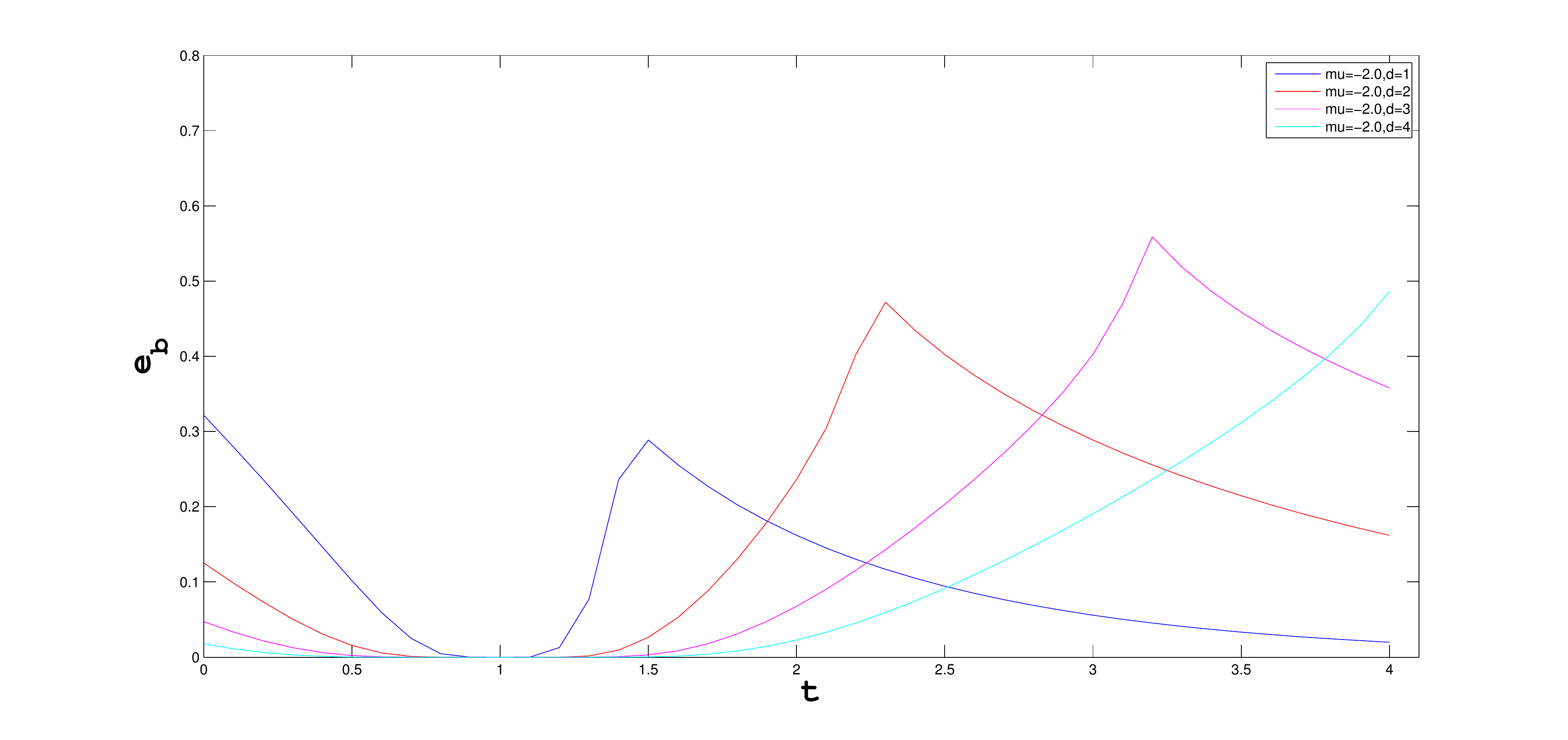}
\caption{ (Color online) Figures show the variation of $\epsilon_b$
with $t$ for different values $\Delta(=d=1,2,3,4)$. The upper panel is for
the $\mu =0$, the middle one is for $\mu=-1.5$ and the lower one
is for $\mu = -2 $.  
 }
\end{figure}
We express the model Hamiltonians of our system in terms of spinless p-wave superconducting
Hamiltonian thus the starting point of our analysis is the seminal paper 
of Kitaev \cite{kitaev}.
\\ 
In the Majorana fermion basis Hamiltonian reduce to:
\begin{eqnarray}
H^{J=1,2,3} = \left( \frac{-{\mu}^J }{2}\right) 
\sum\limits_{n=1}^{N}(1+i\gamma_{B,n}\gamma_{A,n}) 
+ \left(\frac{i}{2}\right)\sum\limits_{n=1}^{N-1}\left[({\Delta}^{J} 
+ {t}^{J} )\gamma_{B,n}\gamma_{A,n+1} 
+ ({\Delta}^{J} -{t}^J)\gamma_{A,n}\gamma_{B,n+1}\right) \nonumber
\end{eqnarray}
Here we use these analytical relations to derive the above Hamiltonian:\\
$ c_{n} = \frac{1}{2} ( {\gamma}_{B,i} + {\gamma}_{A,i} ) $.
${{\gamma}_{B,i}}^{\dagger} = {\gamma}_{B,i} , 
{{\gamma}_{A,i}}^{\dagger} = {\gamma}_{A,i} $. 
$\{ \gamma_{\alpha,x}, \gamma_{\beta,y} \} = 2  \delta_{\alpha,\beta} \delta_{x,y}$. \\
The index $A$ and $B$
are the arbitrary index. In the Kitaev's chain if a Majorana fermion of 
$\gamma_A $ occurs at one end of the chain then the Majorana fermion
$\gamma_B$ must occurs at the other end of the chain \cite{kitaev}.\\ 
Now we discuss the non-topological states of the system.
The first corresponds to $\mu < 0$ but $t=\Delta = 0$.
From this condition, we analyze non-topological states of the three Hamiltonians. \\
For the Hamiltonian, $H_1 $.\\
(1). $ |{\Omega}_a | =0 $ and $ |{\Omega}_b | =0 $, 
(2). $ g_b =0 , g_a =0 $, 
(3). $ |{\Omega}_b | =0 , g_a  =0 $.  
(4). $ |{\Omega}_a | =0 , g_b  =0 $. \\
For the Hamiltonian, $H_2 $, we obtain the same conditions to achieve
the non-topological phase as we obtain in $H_1 $. But the condition for
$\mu$ is different for $H_3$. For $ H_1 $ and $H_2$, $ \delta_1 - 2\beta <0 $,
but for $H_3$ the condition is $ \delta_1 - 2 \beta + 8 \rho -4 < 0$. 
\\ 
Physical explanation of the non-topological phase is the following:\\  
The first term of the above 
Hamiltonian yields a coupling between the Majorana fermion modes
in the same site. In the cavity QED system, this situation corresponds to the
different kind of light-matter interactions which obeys that the coupled  
Majorana fermion modes
condition in each cavity. It is well known that the two Majorana fermion modes
produce a fermion mode. Thus the case of cavity QED system, the non-topological 
state of the system is the fermionic excitations. We consider this state as the
vacuum state of the system where there 
is no gapless Majorana fermion excitation states. We consider this non-topological 
state as a vacuum state of the present system, actually it is the conventional 
superconducting phase.\\
Now our main intention is to find out the topological excitation which appears
as a Majorana fermions. 
The condition for this phase is 
$\mu=0$ and $t=\Delta \neq 0$. Hence the Hamiltonian reduced to
\begin{eqnarray}
H = -it \sum\limits_{n=1}^{N-1} \gamma_{B,n}\gamma_{A,n+1} . \nonumber
\end{eqnarray}
It is very clear from the above Hamiltonian that $ {\gamma_1}= {\gamma_{A,1}} $
and $ {\gamma}_2 = {\gamma}_{B,N} $ are not appear in the Hamiltonian. 
One can also write down the above Hamiltonian by introducing the new
operator $B$. \\
\begin{eqnarray}
B_n = \frac{1}{2} (\gamma_{A,n+1} + i \gamma_{B,n}).\nonumber
\end{eqnarray}
\begin{eqnarray}
H = t \sum\limits_{n=1}^{N-1}(B_{n}^\dagger B_{n} - 1/2).\nonumber
\end{eqnarray}
The ends of the chain has zero energy Majorana fermion modes 
$\gamma_1 = \gamma_{A,1}$ and $\gamma_2 = \gamma_{B,N}$.
These can be considered as a non-local fermion $f=\frac{1}{2} (\gamma_1+i\gamma_2)$. 
This fermion mode is in the zero energy configuration. In this topological phase, 
system is in the doubly degenerate ground state. 
One can understand this by following analysis. \\
If we consider $\left| 0 \right>$ is a ground state then $f\left|0\right>=0$ 
and $\left|{1}\right>=f^\dagger\left|0\right>$ is also a 
ground state with opposite fermion parity. 
The main difference between the conventional superconductor 
and topological superconductor 
is that the  system has a unique ground state with an even parity such a way that 
all the electrons can form cooper pairs. Thus the conventional
superconducting pairing is the vacuum state of the system. Therefore
the non-topological state of the system is in the even parity state and
the doubly degenerate ground state with opposite parity of the system is
the topological state of the system.\\
Here we consider the most general situation, when $\mu \neq 0$ and $t \neq \Delta$.
Here the main intuition is to study the
transition for the topological state of the system to the 
non-topological state of the system.\\
In this arbitrary limit, the Majorana zero modes are no longer $\gamma_{A,1} ,
 \gamma_{B,N}$. In this limit wave function 
decay exponentially into the bulk of the system.
The decay of this wave function results in the 
splitting of the degeneracy between the
states $(f\left|0\right> = 0, \left|1\right> = 
f^\dagger \left|0\right>)$. 
One can write down the effective Hamiltonian as
\beq
H_{eff} = \frac{i}{2} {e_b} b' b'' 
\eeq
when $e_b \propto e^{-\frac{L}{\xi_0}}$, $L $ is the length of the
system. 
 $b'$ and $ b''$ are the Majorana fermion at the left end and
the right end of the chain respectively. $\xi_0^{-1}$ is the smallest of
$|ln|\chi_+|| $ and $|ln|\chi_-||$. When $\xi_0^{-1}$ is $ 0 $
then the coherence length $\xi_0$ is $\infty$.
At this point, the system transit
from topological state to the non-topological state of the matter.
\begin{eqnarray}
\chi_{\pm} = \frac{-\mu \pm \sqrt{\mu^2 - 4t^2 + 4 |\Delta|^2}}{2(t+|\Delta| )} \nonumber
\end{eqnarray}
\beq
{{\xi}_0}^{-1} = Min[ |ln|{\chi}_{+}| |, |ln|{\chi}_{0}| | ]
\eeq
$\epsilon \propto e^-\frac{L}{\xi_0}$ when $\xi_0^{-1} 
\Rightarrow 0$ then $e_b \propto 1$.\\  
In fig. 1, we present the appearance of non-topological 
trivial state for different values of $\Delta (\Delta = 1,2,3,4).$ 
It reveals from our study as we increase the value of $\Delta$, 
the non-topological phase 
occurs for the higher value of t. 
In the present situation the Majorana fermion modes decay 
exponentially into the bulk of the chain. The overlap of these wave 
function result in the splitting in the degeneracy between the 
state $\left|0\right>$ and $\left|1\right>$ by energy 
scale $e^{-\frac{L}{\xi}}$. This figures panel consist of three figures, 
it is clear from these study that as we go away from zero chemical potential
to the higher one the peak at $t=\Delta$ of $e_b $ study gradually decreases.
The coherence length $\xi \rightarrow \infty$ for only $\mu =0 $, where
the system shows the topological to non-topological transition. It is also
clear from the analytical expression for $e_b $ that as we increase the
length of the system keeping the other physical parameters
of the system fixed, the system is in the stable topological state.
Therefore it is clear from our study that the Majorana fermion modes
appear at the edge is more stable for the array of larger length scale
compare to the shorter one.  
Now we explain the corresponding physics in the light of cavity-QED lattice.
In the topological phase, there is no bound between 
Majorana fermion mode between the two ends of the cavity. In the case
of non-topological state, Majorana fermion mode excitation are now bound
at the two end of the cavity. It is also clear from the above analysis of
Kitaev's formula \cite{kitaev} that the that the transition from topological
state to non-topological occurs only for $ \mu =0 $. 
($t \neq 0 \neq \Delta $). But we will also study the topological quantum
phase transition through the variation of winding number calculation in
the next section which yields more new and important result. \\     
\begin{figure}
\includegraphics[scale=0.5,angle=0]{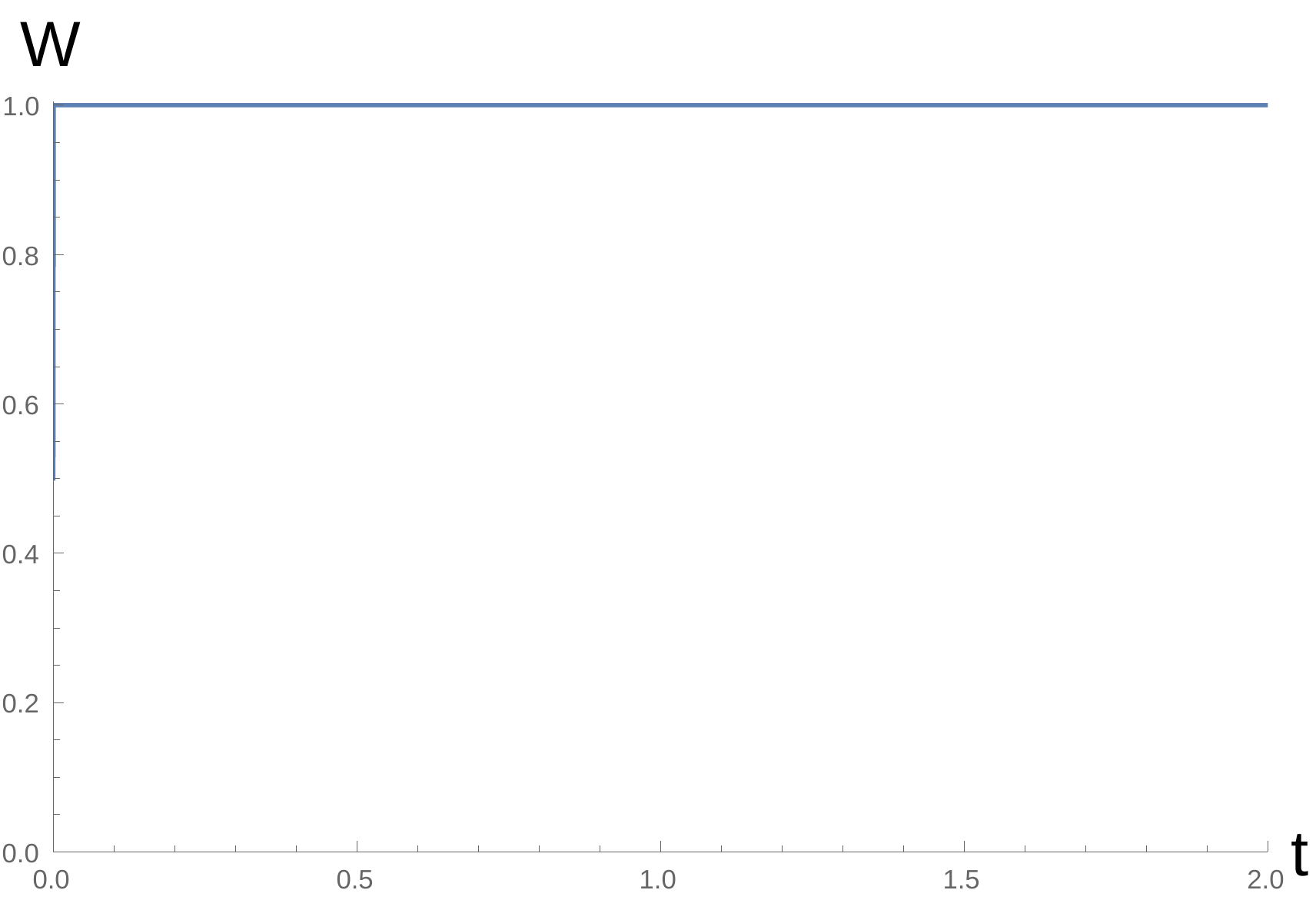}
\includegraphics[scale=0.5,angle=0]{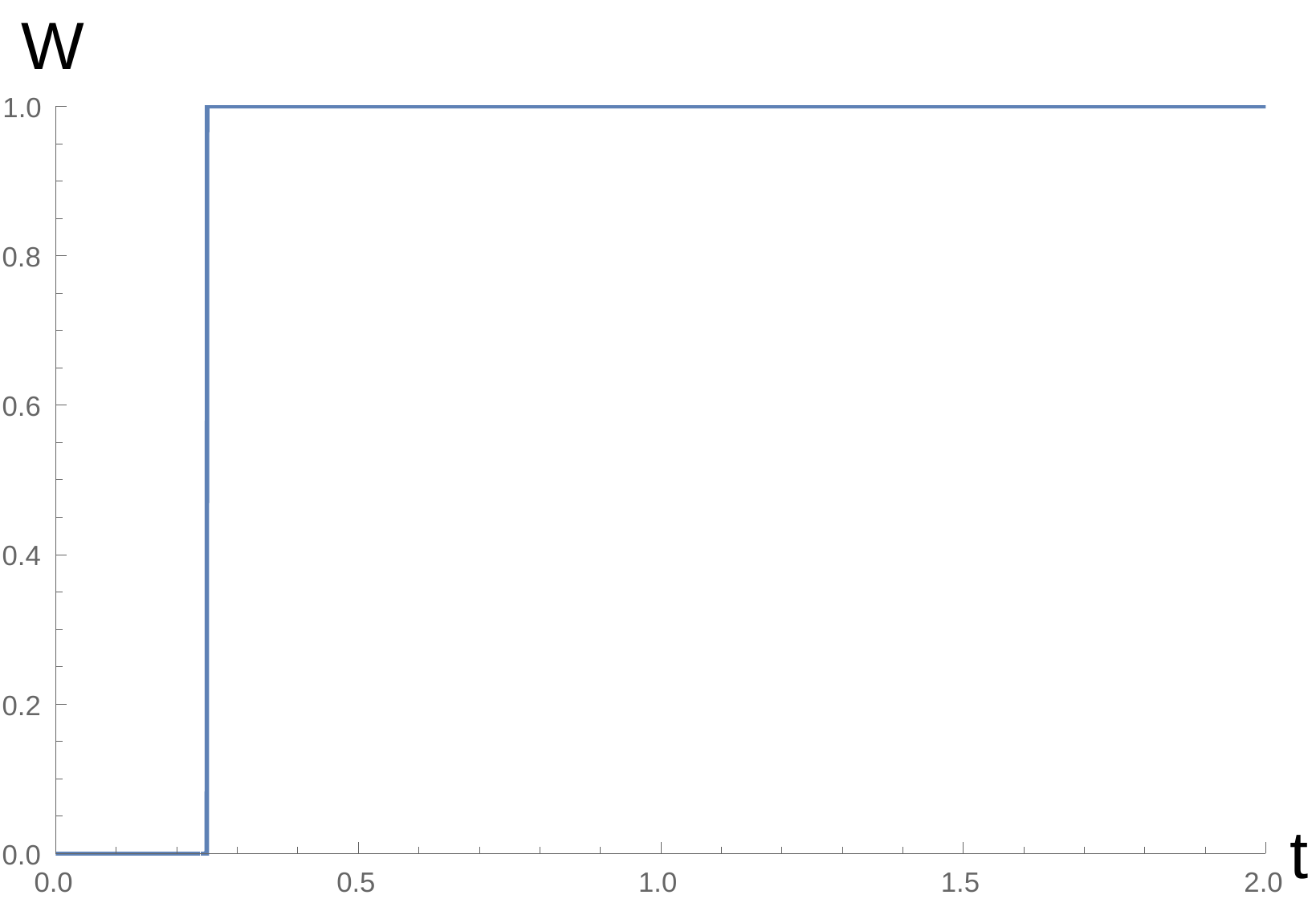}
\includegraphics[scale=0.5,angle=0]{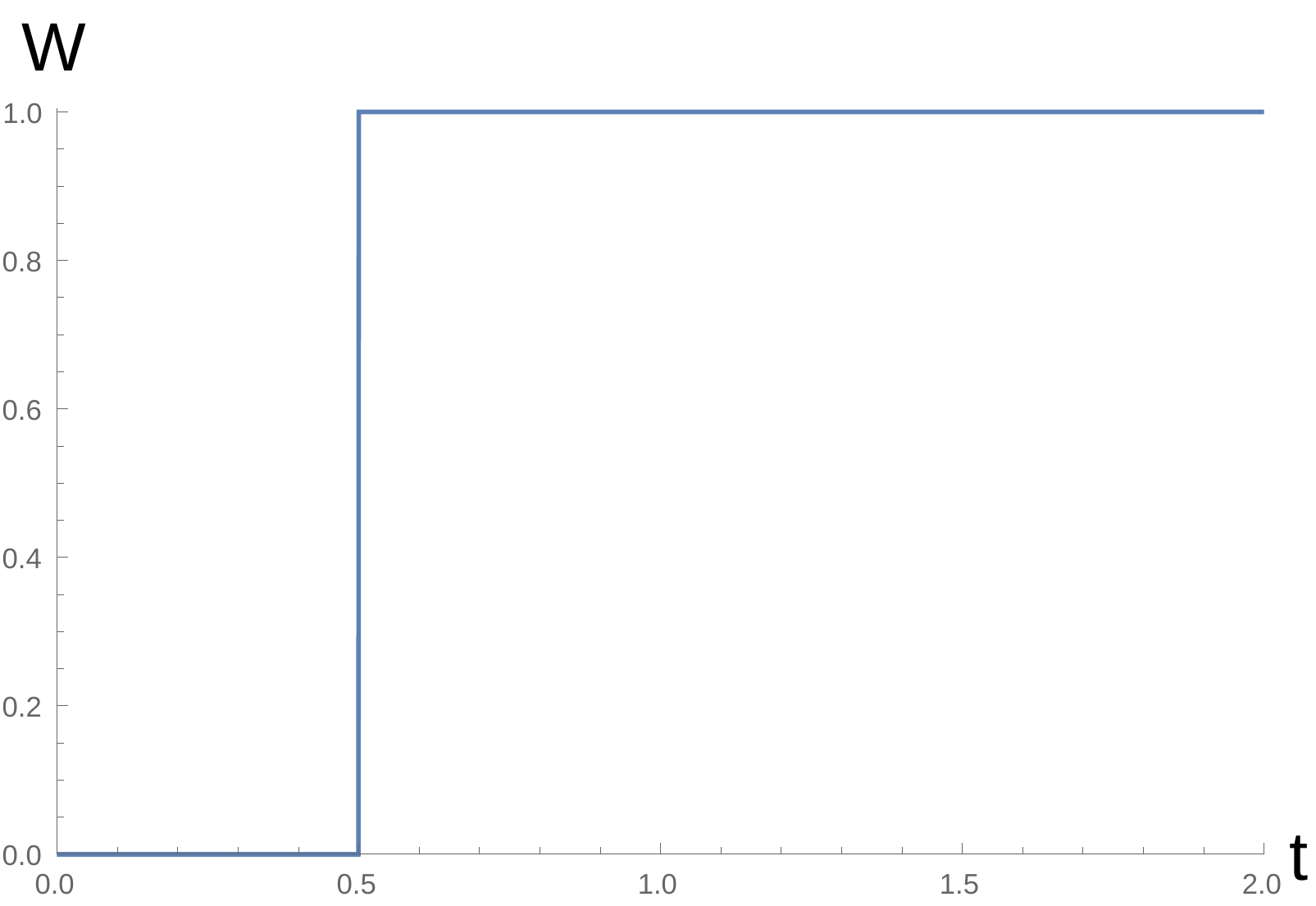}
 \caption{( Color Online), Figures show the variation of winding number
 with t for different values of $\mu$. The upper panel is for the
 $\mu = 0$, the middle panel is for $ \mu = 0.5 $ and the lower
 panel is for $\mu = 1$. Here $\Delta =1$. 
  }
 \end{figure}
\begin{figure}
\includegraphics[scale=0.5,angle=0]{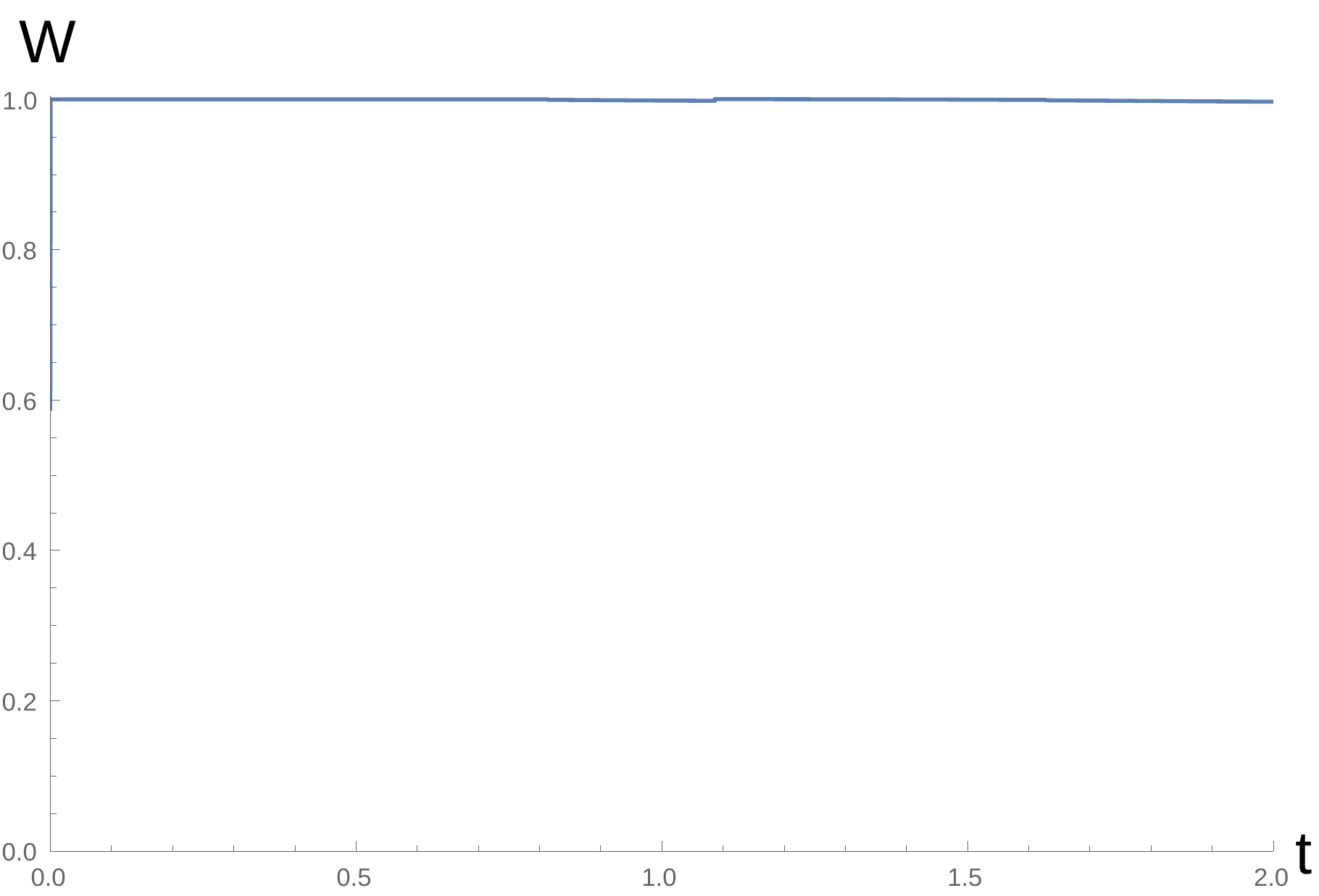}
\includegraphics[scale=0.5,angle=0]{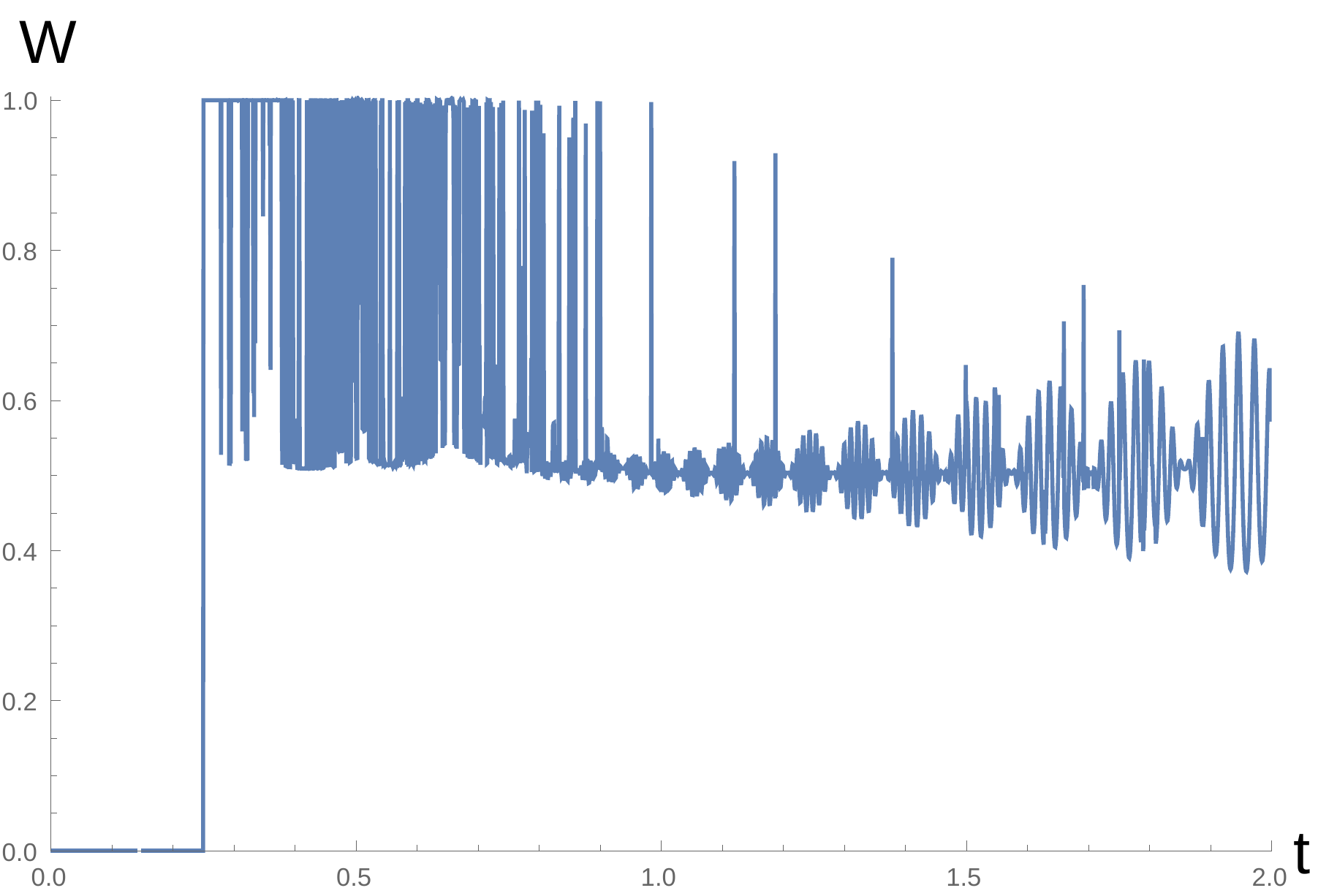}
\includegraphics[scale=0.5,angle=0]{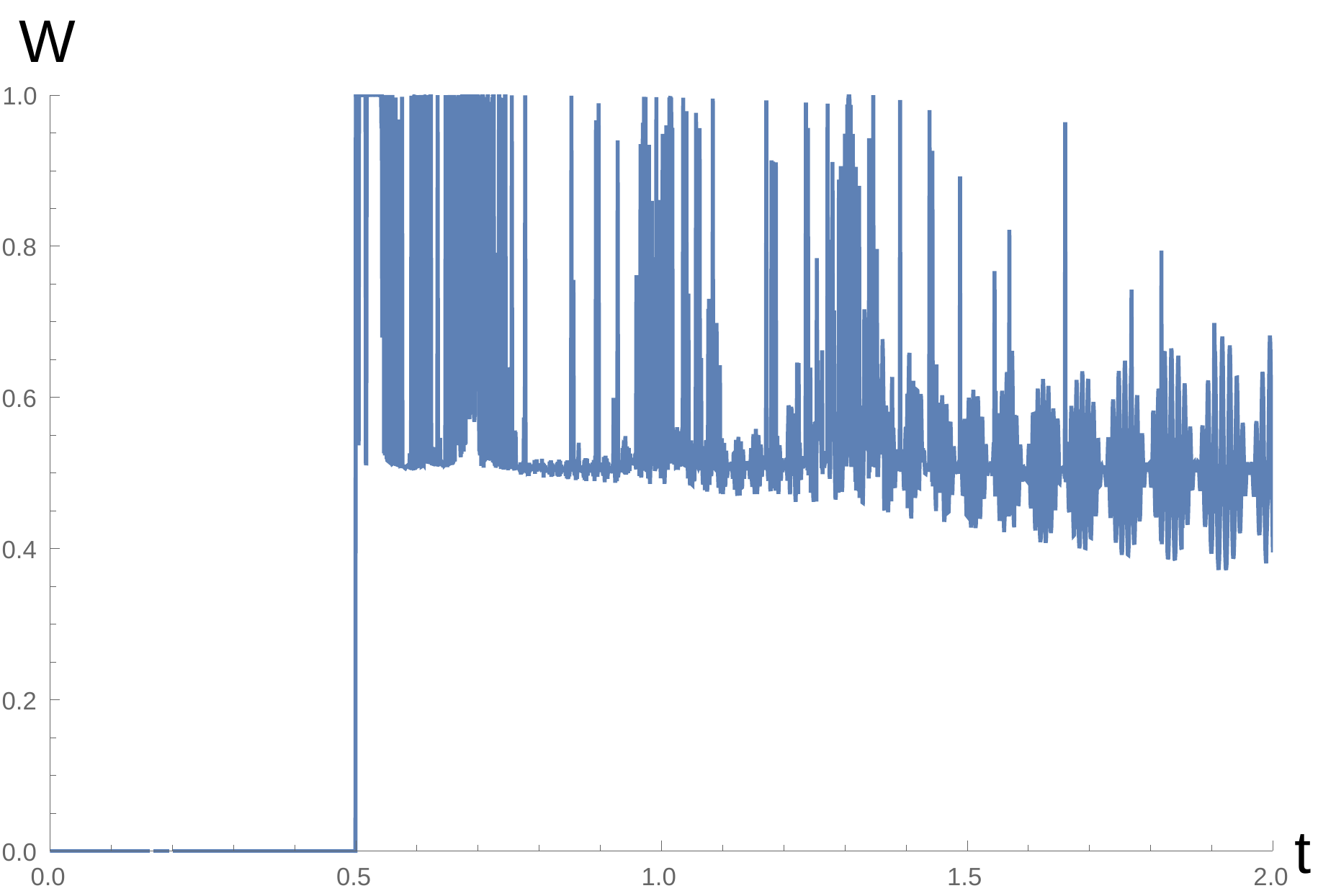}
 \caption{( Color Online), Figures show the variation of winding number
 with t for different values of $\mu$. The upper panel is for the
 $\mu = 0$, the middle panel is for $ \mu = 0.5 $ and the lower
 panel is for $\mu = 1$. Here $\Delta ={10}^{-3} $. 
  }
 \end{figure}
\section{Topological phase transition from the analysis of winding number}
In this section, 
we explicitly discuss the physics of topological to non-topological 
transition from the analysis of winding number calculation. 
We already found the analytical expressions of three Hamiltonians
in momentum space. 
These Hamiltonians are alike to BdG Hamiltonian. 

\begin{eqnarray}
\
   H_{(J= 1,2,3)}(k)=
  \left[ {\begin{array}{cc}
   {{\epsilon}_k}^{( J)} -{\mu}^{( J )} &   i \Delta^{(J)}(k) \\
   -i \Delta^{(J)}(k) & \mu^{(J)}-\epsilon^{(J)}_k \\
  \end{array} } \right] ,
\
\end{eqnarray} 
where $\epsilon_k^{(J)} = 2t^{(J)}cos(k), {\Delta}^J (k) = -2 \Delta^{(J)} sin(k) $. \\
The analytical expression for  
$\epsilon^{(J)}_{k}$ and $\Delta^{J}(k)$ are given in the previous section. \\
Topological phase transition can be ascribed by the 
topological invariant quality. It is convenient to define this 
invariant quantity using the Anderson pseudo-spin approach \cite{anderson}. 
\begin{eqnarray}
{\vec{\chi(k)}}^{J=1,2,3} = {\Delta(k)}^{J} \vec{y} 
+ ({\epsilon_k}^{J} - {\mu}^{J})\vec{z} . 
\end{eqnarray}
One can write the Hamiltonian as $H^{(J=1,2,3)}(k) = {\vec{\chi(k)}}^{J}.\vec{\tau}$ 
where $\vec{\tau}$ are Pauli matrices which act in the particle-hole basis. 
It is very clear from the analytical expression that the pseudo spin define 
in the YZ plane. 
\begin{eqnarray}
\hat{\chi(k)}=\frac{\vec{\chi(k)}}{|\vec{\chi(k)}|}=
cos(\theta_k)\hat{y} + sin(\theta_k)\hat{z} \nonumber \\
\end{eqnarray}
Here the momentum states with periodic boundary condition for a ring $T^{(1)}$ 
and the unit value $\hat{\chi(k)}$ exists on a unit circle $S^{(1)}$ in the YZ plane. 
Therefore $\theta(k)$ is a mapping. $ S^{(1)} \Rightarrow T^1$ and the 
topological invariant is simply the fundamental group of the maping which is 
just the integer winding number.
The physical interpretation of 
this quantity is that the unit vector $\vec{\chi}(k)$ relates in the YZ-plane around 
the Brillouin zone. It is only an integer number and therefore can not vary with 
smooth deformation of the Hamiltonian so large on the quasiparticle gap remains finite. 
At the point of topological phase transition the winding number changes discontinuously.
A topological invariant for $\chi (k)$ is then expressed by \cite{vincent} 
\beq
W = \int \frac{dk}{4 \pi} {\epsilon}_{\alpha \beta} 
\frac{1}{ \hat{\chi}_{\alpha} }  
 \frac{\partial {\hat{\chi}}_{\beta} }{\partial k}. 
\eeq
Here $\alpha$ and $\beta$ are $Y$ and $Z$ two components and 
$\epsilon_{\alpha \beta}$
is anti-symmetric tensor.\\
It is very clear from our study, fig.2, that at $\mu =0$ and $t=0$ system   
shows only the non-topological state of cavity QED lattice. 
Apart from $t=0$, we
observe that the winding number is always one.  
As we go further away from the zero chemical potential, $\mu =0.5, 1$, 
it reveals from our study that
the transition occurs from the non-topological state ($W =0$) to 
topological ($W=1 $) state. 
The non-topological
state persists for a range of $t $ and then follows a transition to the
topological state. 
To the best of our knowledge,
this explicit study of the winding number
calculation is absent in the literature of interacting light-matter system.\\
It is clear from this study that as we go away from $\mu =0 $ the non-topological
state persist for a wider range of $t$. It is related with the following
relations for three different Hamiltonians. \\
For the Hamiltonian $H_2 $, one can write the condition for the persistent
of non-topological state is the following. \\
\beq
{\delta}_1 - 2 \beta = \frac{\gamma_2}{2} 
{(\frac{ {\Omega}_a g_b }{\Delta_a } +  \frac{ {\Omega}_b g_a }{\Delta_b } )}^2 . 
\eeq
For the Hamiltonian $H_3 $, one can write the condition for the persistent
of non-topological state is the following. \\
\beq
{\delta}_1 - 2 \beta + 8 \rho -4 = \frac{\gamma_2}{2} 
{(\frac{ {\Omega}_a g_b }{\Delta_a } +  \frac{ {\Omega}_b g_a }{\Delta_b } )}^2 . 
\eeq
In fig.3, we study the behavior of winding number with t for smaller
value of $\Delta = {10}^{-3} $. It is clear to us from this study
that the system is in the topological state for the zero chemical potential.
As we away from the zero chemical potential the system is in the
non-topological state, i.e. there is no sharp topological phase transition
in the system. Fluctuation 
of the winding number is extremely large that one cannot predict about 
the definite topological phase transition. It can be understand in the
following way, as $\Delta \rightarrow 0$, the system is simple a fermionic
chain with out any spinless p-wave superconductivity in the Hamiltonian.
Therefore the bulk gap of the system is absent which implies that system
has no topological state with two Majorana fermion mode at the edge.\\   
 
Acknowledgement: The author would like to acknowledge the several interesting 
and important discussions during the 
international school
and discussion meeting on ``Topological State of Matter" at Harishchandra Research
Institute at Allahabad. The author would like to acknowledge the DST project,
Govt. of India. Finally, author would like to acknowledge the library of Raman
Research Institute for extensive help.\\

\section{Appendix}

\bea
\beta  & = &  \frac{1}{2} [\frac{{|{\Omega_b}|}^2 }{4 {\Delta}_b }
({\Delta}_b - \frac{{|{\Omega_b}|}^2 }{4 {\Delta}_b } -  
 \frac{{|{\Omega_b}|}^2 }{4 ( {\Delta}_a  - {\Delta}_b )} - {\gamma_b} {g_b}^2 
- {\gamma_1} {g_a}^2 \non\\ 
& & + {\gamma_1}^2 \frac{{g_a}^4 }{{\Delta_b}} - (a \leftrightarrow b)] 
\eea

$ \gamma_{a,b} = \frac{1}{N} \sum_{k} \frac{1}{ {\omega}_{a,b} - {\omega}_k } $
$ \gamma_{1} = \frac{1}{N} \sum_{k} \frac{1}{ ( {\omega}_{a}+ 
 {\omega}_{b})/2 - {\omega}_k } $ and
$ \gamma_{2} = \frac{1}{N} \sum_{k} \frac{e^{ik} }{ ( {\omega}_{a}+  {\omega}_{b})/2 
- {\omega}_k } $
${\delta_1} = {\omega}_{ab} - ({\omega}_a - {\omega}_b )/2 $, 
${\Delta}_a = {\omega}_e - {\omega}_a$.  
${\Delta}_b = {\omega}_e - {\omega}_a -({\omega}_{ab} - {\delta_1})$.
${{\delta}_a}^{k} = {\omega}_e - {\omega}_k $,
${{\delta}_b}^{k} = {\omega}_e - {\omega}_k  -({\omega}_{ab} - {\delta_1}) $,
${\omega}_k = {\omega}_c + J_c \sum_{k} cosk $. \\

$ {\beta_2} = \sum_{j= a,b}  \frac{{|{\Omega_j}|}^2 }{4 ( {\Delta}_j  - \tilde{{\Delta}_b} )}.
4 \tilde{\gamma_{jb}} {g_j}^2  $ \\
\bea
\beta_3  & = &   [\frac{{|{\Omega_b}|}^2 }{16 { {\Delta}_b}^2 }
(4 {\Delta}_b - \frac{{|{\Omega_a}|}^2 }{4 {\Delta}_b } - \non\\
& & \frac{{|{\Omega_b}|}^2 }{4 ( {\Delta}_a  - {\Delta}_b )} - 
\frac{{|{\Omega_b}|}^2}{\Delta_b}
- \sum_{j= a,b}  \frac{{|{\Lambda_j}|}^2 }{4 ( {\Delta}_j  - \tilde{{\Delta}_b} )}.
4 {\gamma_{jb}} {g_j}^2 )  
+ {\gamma_{bb}}^2 \frac{{g_b}^4 }{{\Delta_b}} - (a \leftrightarrow b)]
\eea
Here ${\gamma}_1 = \frac{1}{N} \sum_{k} \frac{1}{\omega - {\omega}_k }$,
${\gamma}_2 = \frac{1}{N} \sum_{k} \frac{e^{ik}}{\omega - {\omega}_k }$,
$ {\gamma}_{aa}  = {\gamma}_{bb} = 
\frac{1}{N} \sum_k \frac{1}{\omega - \omega_k} $.\\

Here $J_z =0 $, this condition, yields the following analytical relation
between the different physical parameters of the system.\\
\beq
\frac{g_b \Delta_a}{g_a \Delta_b} = \frac{\Omega_a}{\Omega_b} .
\eeq
And also from the condition of $J_y =0 $, i.e., $J_1 = J_2 $. 
\beq
\frac{g_a \Delta_a}{g_b \Delta_b} = \frac{\Omega_a}{\Omega_b} .
\eeq
From the above equation, We get the following relations, $ g_b = g_a $ to get
the transverse Ising model. \\
Derivation of Eq.26:\\
Following Ref. \cite{kote,sig} , one 
can write the order parameter as the sum of singlet ($\psi (k)$)
and triplet ($P (k) $). The singlet and triplet component satisfy the following
relation $ {\psi} (k) = {\psi} (-k)$ and $ P (k) = - P (-k) $. One can write the
general expression for order parameter as
\beq
\vec{\Delta} (k) = i {\psi} (k) {\sigma}_y + i P (k) {\sigma}_y \sigma
\eeq

Therefore for the p-wave superconductor, one can write the mean field Hamiltonian
as
\beq
H = \int dk [ \sum_{\alpha} {\epsilon}_{\alpha} (k) {c_{\alpha} (k)}^{\dagger} c_{\alpha} (k)
+ (i {c_{\alpha}} (k) P(k). (\sigma \sigma_y )_{\alpha \beta} [{ c_{\beta}}^{\dagger} + h.c ]  
\eeq
The first part of the Hamiltonian is the single particle energy and the second part
is the pairing energy. Here we consider, $ P(k) = (0, -{\Delta} (k), 0)$. Finally
the Hamiltonian become 
\beq
H = \int dk [ \sum_{\alpha} {\epsilon}_{\alpha} (k) {c_{\alpha} (k)}^{\dagger} c_{\alpha} (k)
- \frac{1}{2} \sum_{\alpha} 
[ 2 i \Delta (k) {c_{\sigma}}^{\dagger} (k) {c_{\sigma}}^{\dagger}
(-k)   
-  2 i \Delta (k) {c_{\sigma}} (k) {c_{\sigma}}
(-k)   
\eeq
In the above expression, we use ${\Delta} (k) = - {\Delta} (-k) $.
Finally we can write this hamiltonian in Nambu representation
\beq
H = \frac{1}{2} \int dk \sum_{\alpha} {\psi_{\alpha}}^{\dagger} H_k {\psi}_{\alpha} (k)
\eeq
$\psi_{\alpha} (k) = ( c_{\alpha} (k), - {c_{\alpha}}^{\dagger}) $. The half factor is
to balance the double counting.
$ {H}_{k} = \left (\begin{array}{cc}
      {\epsilon}_{\alpha} (k)  & - 2 i {\Delta}_{\alpha} (k) \\
   2 i {\Delta}_{\alpha} (k)   & - {\epsilon}_{\alpha} (k) 
        \end{array} \right ) $


\begin{references}

\bib{berni} Anderi Bernivig and Taylor L. Hughes, "Topological Insulator
and Topological Superconductor", (Princeton University Press, Princeton, 2013).

\bibitem{ss} Subir Sachdev, {\it Quantum Phase Transitions}, 
(Cambridge University Press, 2001 ).

\bibitem{nayak} C. Nayak $et~al.$, Rev. Mod. Phys {\bf 80}, 1083 (2008).

\bib{horo} S. Horoche  and J. M. Raimond  2006 in {\it Exploring the Quantum Atoms,
Cavities, and Photons}, (Oxford University Press).

\bib{agarwal} G. Agarwal, "Quantum Optics", (Cambridge University Press,
Delhi 2013).

\bibitem{kitaev} A. Y. Kitaev, Physics-Uspekhi {\bf 44}, 131 (2001).

\bib{hart1} Hartmann Michael J, Fernando G S,  Brando L and 
Plenio Martin B 2006
Nature Phys {\bf 462} 849;
Hartmann Michael J, Fernando G S, Brando L and Plenio Martin B 2008, Laser and 
Photonics Rev. {\bf 2} 527. 

\bib{hart2} Hartmann Michael J, Fernando G S, Brando L and Plenio Martin B 2007,
Phys. Rev. Lett {\bf 99} 160501.

\bib{sujop} S. Sarkar, arXiv/con-mat:1309.7742 .

\bibitem{majo} E. Majorana, Nuovo Cimento {\bf 14}, 171 (1937).

\bibitem{wil} F. Wilczek, Majorana returns, Nature Physics {\bf 5}, 614

\bib{kote} P. Kotetes, "Topological Insulator and Superconductors"-
Notes of TKMI 2013/2014.

\bib{sig} M. Sigrist and K. Ueda, Rev. Mod. Phys {\bf 63}, 239 (1991).

\bib{bardyn} C. E. Bardyn and A. Imamoglu, Phys. Rev. Lett {\bf 109},
253606 (2012).

\bib{anderson} P. W. Anderson, Phys. Rev {\bf 110}, 827 (1958).

\bib{vincent} W. Vincent Liu, "Selected Topics in Modern Many-Body
Theory" in Summer School of Department of Physics at Tsinghua, University
of Beijing, China (2013). 

\end{references}
\end{document}